\documentclass[12pt]{article}
\usepackage{graphicx}
\usepackage{cite}
\date{}
\begin{document}
\title{New result on the measurement of the direct photon 
emission in $K^+ \rightarrow \pi^+ \pi^0 \gamma$ decay
~\\
~\\
{\normalsize M.A.~Aliev$^{a}$\thanks{Corresponding 
author.\emph{E-mail address:} malik@inr.ru},
Y.~Asano$^b$, 
P.~Depommier$^c$, 
M.~Hasinoff$^d$,\\
K.~Horie$^e$,
Y.~Igarashi$^f$,
J.~Imazato$^f$, 
A.P.~Ivashkin$^a$,\\
M.M.~Khabibullin$^a$,
A.N.~Khotjantsev$^a$, 
Y.G.~Kudenko$^a$,
G.Y.~Lim$^f$, \\
O.V.~Mineev$^a$, 
J.A.~Macdonald$^g$\thanks{Deceased},
C.~Rangacharyulu$^h$,
S.~Sawada$^f$, 
S.~Shimizu$^i$ \\} 
~\\
{\normalsize (The E470 KEK-PS Collaboration)
~\\
~\\
$^a$ Institute for Nuclear Research RAS, Moscow, 117312~Russia \\
$^b$ Institute of Applied Physics, University of Tsukuba, Ibaraki 305-0006, Japan \\
$^c$ Laboratoire de Physique Nucleaire, Universite de Montreal, Montreal, Quebec, Canada~H3C~3J7 \\
$^d$ Department of Physics and Astronomy, University of British Columbia, Vancouver, Canada~V6T~1Z1 \\
$^e$ Research Center for Nuclear Physics, Osaka University, Ibaraki, Osaka 567-0047, Japan \\
$^f$ Institute of Particle and Nuclear Studies (IPNS), High Energy Accelerator Research Organization
(KEK), Ibaraki 305-0801, Japan \\
$^g$ TRIUMF, Vancouver,  British Columbia, Canada V6T 2A3 \\
$^h$ Department of Physics, University of Saskatchewan, Saskatoon, Canada S7N 5E2 \\
$^i$ Department of Physics, Osaka University, Osaka 560-0043, Japan \\
    }}
\maketitle
\begin{abstract}
We present a new result on the $K^+\rightarrow\pi^+\pi^0\gamma$ decay measurement 
using stopped kaons. The best fit to the decay spectrum comprised of 10k events 
gives a branching ratio for the direct photon emission of
$
[3.8\pm0.8(stat)\pm0.7(syst)]\times10^{-6}
$
in the $\pi^+$ kinetic energy region of 55 to 90 MeV. This result has been
obtained with the assumption that there is no component due to interference 
with the inner bremsstrahlung.
\end{abstract}

\newpage
\section{Introduction}

The chiral anomaly~\cite{adler,Adler:2004ih} is a fundamental property of chiral 
quantum field theories such as the Standard Model of quantum chromodynamics (QCD). 
Its theoretical origin and 
mathematical properties are well understood, while experimental tests, which are 
crucial for the theoretical basis of particle physics, are relatively rare. 
Although the chiral anomaly can be interpreted as a short-distant effect, it 
manifests itself most directly in the low-energy interactions of the pseudoscalar 
mesons~\cite{weinberg,gasser1,gasser2,wess}.
The chiral anomaly manifests itself also in the non-leptonic weak interactions. 
As already shown in Refs.~\cite{ecker1,bijnen}, only radiative K decays (among 
them $K^+\rightarrow \pi^+ \pi^0 \gamma$) are sensitive to this anomaly.
 
The total amplitude of the $K^+ \rightarrow \pi^+ \pi^0 \gamma$ decay can be 
written as a sum of inner bremsstrahlung (IB) and direct emission (DE) amplitudes.
The IB amplitude is associated with the $K^+ \rightarrow \pi^+ \pi^0$ ($K_{\pi 2}$)
decay, in which the photon is emitted from the outgoing $\pi^+$, and is completely
predicted by the quantum electrodynamics in terms of the $K_{\pi 2}$ amplitude~\cite{Low}.
The DE amplitude, in which the photon is emitted directly from one of the 
intermediate states of the decay, can in turn be decomposed into electric and 
magnetic parts. 

Within the framework of Chiral Perturbation Theory (ChPT), the magnetic amplitude 
appears first at $O(p^4)$. There are two different ways for its manifestation: 
the reducible amplitude~\cite{ecker1}, which can be derived directly from the 
Wess-Zumino-Witten functional~\cite{wess}, and the direct contributions~\cite{bijnen,cheng}, 
which are subject to some theoretical uncertainties. In both cases the direct 
emission amplitudes are dominated by the anomaly. For the electric amplitude there 
is no definite prediction from ChPT. The experimental study of the electric 
amplitude, which could be observed through an interference (INT) with the IB 
amplitude, is of high interest not only for ChPT but also for possible 
non-standard-model effects such as the CP-violation asymmetry between the 
$K^+\rightarrow\pi^+\pi^0\gamma$ and the 
$K^-\rightarrow\pi^-\pi^0\gamma$ decay widths.

The remarkable feature of the $K^+ \rightarrow \pi^+ \pi^0\gamma$ and 
the $K_L \rightarrow \pi^+ \pi^-\gamma$, is that the normally dominant
IB amplitude is strongly suppressed because it violates the $\Delta I$=1/2 
rule. Therefore the DE amplitude is relatively enhanced. The DE component 
can also be isolated kinematically. These features simplify the experimental
verification of the anomalous amplitude, making the $K^+ \rightarrow \pi^+ \pi^0 \gamma$
decay an important instrument to investigate the low energy structure of QCD.
 
A conventional way to express the differential decay rate of the $K^+ \rightarrow \pi^+
\pi^0 \gamma$ decay is to use terms of $T_+$ and $W$, where $T_+$ is the kinetic energy
of the $\pi^+$ in the $K^+$ rest frame, and $W$ is an invariant parameter defined as
$W=(P\cdot q)(p_+\cdot q)/(m^2_{\pi^+}M^2_{K^+})$. Here, $P$, $p_+$ and $q$ are the
4-momenta of the $K^+$, $\pi^+$ and $\gamma$, respectively. The differential decay rate
can be written in terms of the IB component rate as~\cite{DAmbr_Isid}
\begin{eqnarray}
\frac{\partial^2 \Gamma}{\partial T_+\partial W}=\frac{\partial^2 \Gamma_{IB}}{\partial
T_+\partial W} \left\{1+2\frac{m^2_{\pi^+}}{M_{K^+}}Re \biggl(\frac{E}{eA}
\biggr)W^2+\frac{m^4_{\pi^+}}{M^2_{K^+}} \biggl(\left|\frac{E}{eA}
\right|^2+\left|\frac{M}{eA} \right|^2 \biggr)W^4 \right\},
\end{eqnarray}
where $A$ is the decay amplitude for the $K^+\rightarrow \pi^+\pi^0$, and $E$ and $M$ are 
the electric and magnetic amplitudes of the DE component, respectively.

The experimental status of the measurement of the DE component of the $K^+ \rightarrow
\pi^+ \pi^0 \gamma$ decay is shown in Table~\ref{tab:exp_situation}. 
\begin{table*}
\begin{center}
\caption{Summary of the previous measurements of the direct photon
emission in $K^{\pm}\rightarrow \pi^{\pm}\pi^0 \gamma$ decays.}
\begin{tabular}{|c|c|c|c|c|}
\hline
                                    & Sort of     &           & Number  &    \\
Experiment                          &  experiment & Kaon      & of events    & $Br(DE)$   \\
\hline
BNL \ \ \ \ \ \ 1972~\cite{Abram} & in-flight & $K^{\pm}$ & 2100         & $[1.56 \pm 0.35 \pm 0.5]\times 10^{-5}$  \\
\hline
CERN  \ \ \ \ \ 1976~\cite{Smith} & in-flight & $K^{\pm}$ & 2461         & $[2.3 \pm 3.2 ]\times 10^{-5}$          \\
\hline
Protvino \ \ 1987~\cite{Bolot} & in-flight & $K^-$     & 140          & $[2.05 \pm 0.46^{+0.39}_{-0.23}]\times 10^{-5}$ \\
\hline
BNL E787 2000 \cite{Adler}         & stopped   & $K^+ $    & $2 \times 10^4$ & $[4.7 \pm 0.8 \pm 0.3]\times 10^{-6}$ \\
\hline
KEK E470 2003~\cite{Shim2}         & stopped   & $K^+ $    & 4434         & $[3.2 \pm 1.3 \pm 1.0]\times 10^{-6}$ \\
\hline
Protvino \ \ 2004~\cite{Uvar}      & in-flight & $K^-$     & 930          & $[3.7 \pm 3.9 \pm 1.0]\times 10^{-6}$ \\
\hline
\end{tabular}
\label{tab:exp_situation}
\end{center}
\end{table*}
As one can see,
there is a large discrepancy between the results of the first three~\cite{Abram,Smith,Bolot} 
and the second three~\cite{Adler,Shim2,Uvar} experiments.
These results can also be compared with corresponding theoretical predictions for the 
branching ratio of the DE component in the same kinetic energy region of the 
$\pi^+$~\cite{ecker1,bijnen,cheng}.
The above-mentioned situation encouraged us to perform a new analysis of the experimental
data of the E470 experiment.

\section{Experiment}

\begin{figure}[h]
\begin{center}
\includegraphics[width=11cm]{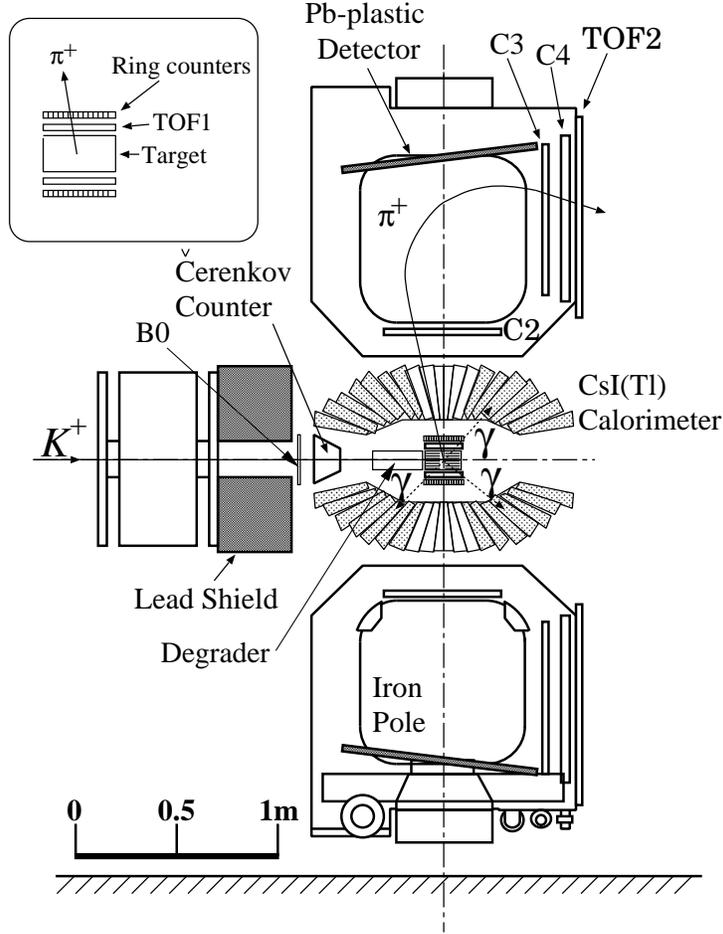}
\end{center}
\caption{The layout of the detector for E470 experiment.}
\label{fig:setup}
\end{figure}

Experiment E470 was performed at the KEK 12 GeV proton synchrotron using the E246
experimental apparatus constructed to search for T-violating transverse muon
polarization in $K^+\rightarrow \pi^0\mu^+\nu$ ($K_{\mu 3}$) decay~\cite{Abe}. 
In addition to the $T$-violation search, spectroscopic studies for various decay 
channels have also been successfully performed using this detector system~\cite{ShLHSh}.
The apparatus is shown in Fig.~\ref{fig:setup}, 
and is described in detail elsewhere~\cite{Grig,Ivash,Demen,Macd}. A separated 
$K^+$ beam ($\pi^+/K^+\sim 7$) of 660 MeV/c was used with a typical intensity of 
$1.5 \times 10^5$ kaons per spill of 2~s duration with a 4~s repetition time. 
To distinguish $K^+$'s from $\pi^+$'s a Cherenkov counter~\cite{Grig} was used.
A beam counter B0~\cite{Macd}, an assembly of 22 plastic scintillating counters, 
placed before the Cherenkov counter, was used to obtain the $K^+$ and $\pi^+$ 
beam profiles. Kaons were slowed down in an Al+BeO degrader, and stopped in an 
active target made of 256 scintillating fibers, which was located at the centre 
of a 12-sector superconducting toroidal spectrometer. The charged $\pi^+$ from 
kaon decay passed through one of the twelve fiducial (also called TOF1) counters, 
surrounding the target, and entered one of the spectrometer sector gaps, and was 
then detected by a TOF2 counter, located at the exit of the spectrometer. 
Tracking of charged particles and their momentum analysis was done using multi-wire 
proportional chambers at the entrance (C2) and exit (C3 and C4) of each magnet 
sector, as well as by the active target and an array of ring counters~\cite{Ivash} 
surrounding the target. The energies and angles of the photons were measured by 
the CsI(Tl) photon calorimeter consisting of 768 CsI(Tl) modules~\cite{Demen}. 
The CsI(Tl) barrel had twelve holes for the charged particle entry to the toroidal 
spectrometer, and covered a solid angle of about 0.75$\times 4 \pi$~sr.

To collect the $K^+ \rightarrow \pi^+ \pi^0 \gamma$ data the following trigger was used
\[
C_K \times [Fid_i \times TOF2_i] \times 3\gamma \times (1,2)B0 \times (1,2)ring,
\]
where $C_K$ is a kaon Cherenkov counter, $Fid_i$ is a hit in the $i$th fiducial (TOF1)
counter or in its immediate neighbors, $TOF2_i$ is a hit in the $i$th $TOF2$ counter,
$3\gamma$ is 3-photon clusters in CsI(Tl) barrel, and $(1,2)B0$ and $(1,2)ring$ are
multiplicities of the B0 counter and ring counters $\le$2, respectively.

\section{Analysis}

In this analysis we aimed to improve our previous result~\cite{Shim2} of the 
measurement of the direct photon emission in $K^+ \rightarrow \pi^+ \pi^0 \gamma$ 
decay by reanalyzing the experimental data accumulated in 2001. For this 
purpose a new analysis strategy was adopted. This included: 
1)~selecting as many good $K^+ \rightarrow \pi^+ \pi^0 \gamma$ events as possible 
by imposing several loose cuts; 
2)~increasing the quality of data by improving 
the pairing efficiency of the photons from $\pi^0$ decay, and 
3)~improving the $K_{\pi3}$ background rejection. 
The analysis strategy was carefully studied and optimized by a Monte Carlo simulation.

The decay time of the stopped $K^+$ in the target was defined by a signal in 
the TOF1 counter. To reject in-flight $K^+$ decays this time was required to 
be more than 3 ns later than the $K^+$ arrival time into the target.

For the analysis only events with three photon clusters in the CsI(Tl)
calorimeter have been selected. This condition considerably suppresses 
the backgrounds from the $K_{\pi2}$, $K_{e3}$, $K_{\mu3}$ and $K_{\pi3}$ 
modes, although some fraction of these events pass the trigger due to 
different reasons-- the $K_{e3}$ and $K_{\mu3}$ modes due to an accidental 
photon hit in the CsI(Tl) calorimeter; the $K_{\pi3}$ mode due
to the escape of one photon into a CsI(Tl) calorimeter hole. The $K_{\pi2}$ 
events pass through the trigger condition mostly due to identification 
one of the two photons from  $\pi^0$ decay dissipated in two parts as two 
photons. As a result of such misidentification we have three photon 
clusters in the CsI(Tl) calorimeter instead of two.

To separate $e^+$'s and $\mu^+$'s from $\pi^+$'s, the squared mass of the
charged particles $M^2_{TOF}$ was calculated using the time of flight of 
the charged particles from the TOF1 to the TOF2 counter, and their 
specific energy deposition $dE/dx$ in the TOF2. Such separation completely 
remove the remaining $K_{e3}$ background, and considerably reduced the 
$K_{\mu3}$ background.

Constraints on the reconstructed $K^+$ mass $M_{K^+}=E_{\pi^+}+\sum_{i=1}^3E_{\gamma i}$
were set to be $420<M_K<510$ MeV/c$^2$. To set a cut on the momentum
$\vec{p}_K=\vec{p}_{\pi^+}+\sum_{i=1}^3 \vec{p}_{\gamma_i}$ we used the
parameter $\Delta p=\sqrt{{p_x}^2+{p_y}^2+{p_z}^2}$, where 
$p_{\alpha}=p_{\pi^+\alpha}+\sum_{i=1}^3p_{\gamma_i \alpha}$ 
($\alpha=x, y, z$), and used the condition of $\Delta p < 60$~MeV/c.

To reject in-flight decays of $\pi^+$'s and suppress the scattering of
charged particles from the magnetic pole faces, a track consistency cut 
$\chi^2<10$ has been imposed. The remaining $K_{\pi2}$ background has 
been removed by requiring the $\pi^+$ momentum to be less than 180~MeV/c.

For further analysis we need to reconstruct the $\pi^0$. Since we have three 
photons in the CsI(Tl), there are also three possible combinations to form 
the $\pi^0$. To improve the pairing efficiency of the photons the following 
method was developed. The quantities $Q^2_{i,j}$ ($i$, $j$=1,2,3), 
sums of six asymmetric terms for each pair of photons comprised of the $i$th 
and $j$th photon, were introduced as follows:
\[
Q_{i,j}^2 = \frac{(\theta^{exp}_{\pi^0 \gamma_{6-i-j}}-\theta_{\pi^0
\gamma})^2}{(\sigma_{\theta_{\pi^0 \gamma}})^2} +
\frac{(\theta^{exp}_{\pi^+ \gamma_{6-i-j}}-\theta_{\pi^+
\gamma})^2}{(\sigma_{\theta_{\pi^+ \gamma}})^2}
+ \frac{(E^{calc 0}_{\pi^0(\gamma_i \gamma_j)}-E^{calc 1}_{\pi^0(\gamma_i
\gamma_j)}-\Delta E_{\pi^0})^2}{(\sigma_{E_{\pi^0}})^2} +
\]
\begin{equation}
\label{eq:Q_i_j}
\frac{(m_{\gamma_i\gamma_j}-M_{\gamma\gamma})^2}{(\sigma_{m_{\gamma\gamma}})^2}
+ \frac{(E_{\gamma_{6-i-j}}^{calc}-E_{\gamma_{6-i-j}}^{exp}-\Delta
E_{\gamma})^2}{(\sigma_{E\gamma})^2}+ \frac{(\theta^{exp}_{\pi^+
\pi^0_{i,j}}-\theta_{\pi^+\pi^0})^2}{(\sigma_{\theta_{\pi^+ \pi^0}})^2}.
\end{equation}
Here, the superscripts "exp" and "calc" stand for the experimental and
calculated values, respectively. \ So, $E^{calc 0}_{\pi^0(\gamma_i \gamma_j)}$ 
and $E^{calc1}_{\pi^0(\gamma_i\gamma_j)}$ are the calculated energies of 
the $\pi^0$ comprised of a combination of $i$th and $j$th photons, using 
the formulae:
\begin{equation}
E^{calc 0}_{\pi^0(\gamma_i\gamma_j)}=\frac{\sqrt{2} m_{\pi^0}}
{\sqrt{(1-\cos{\theta^{exp}_{\gamma i,\gamma j}})
\biggl[1-\displaystyle\biggl(\frac{\mathstrut
(E^{exp}_{\gamma_i}-E^{exp}_{\gamma_j})}{(E^{exp}_{\gamma_i}+E^{exp}_{\gamma_j})}\biggr)^2}\biggr]},
\end{equation}
\begin{equation}
E^{calc1}_{\pi^0(\gamma_i\gamma_j)}=\frac{M_{K^+}^2+m_{\pi^0}^2-m_{\pi^+}^2}{2
M_{K^+}}-\frac{2 E^{exp}_{\gamma_{6-i-j}} (E^{exp}_{\pi^+}-P^{exp}_{\pi^+}
\cos{\theta^{exp}_{\pi^+ \gamma_{6-i-j}}} )}{2 M_{K^+}}.
\end{equation}
$E^{calc}_{\gamma_{6-i-j}}$, the energy of the $(6-i-j)$th photon
(supposing it is the unpaired one), is calculated from the following formula:
\begin{equation}
E^{calc}_{\gamma_{6-i-j}}=\frac{(M^2_K+m^2_{\pi^+}-m^2_{\pi^0}-2 M_K E_{\pi^+}^{exp})}
     {2(M_K-E_{\pi^+}^{exp}+p_{\pi^+}^{exp}\cos\theta_{\pi^+ \gamma_{6-i-j}}^{exp})}.
\end{equation}
$
m_{\gamma_i \gamma_j}= \sqrt{2 E^{exp}_i E^{exp}_j(1-\cos\theta^{exp}_{\gamma_i\gamma_j})}
$
is the invariant mass of the $(i,j)$ photon combination. Other parameters in
the $Q_{i,j}^2$ expressions, which were obtained using MC simulation, are given in
Table~\ref{tab:Q_param}.
\begin{table*}
\begin{center}
\caption{The values of parameters in the $Q^2_{i,j}$ expressions.}
\begin{tabular}{|c|c|c|c|}
\hline
  &  &  & \\
Expression  & Offset & $\sigma$ (Expression $<$0)  & $\sigma$ (Expression $>$0)\\
\hline
$E_{\gamma_{6-i-j}}^{calc}- E_{\gamma_{6-i-j}}^{exp}-\Delta E_{\gamma}$, MeV & 
$\Delta E_{\gamma}= 2.54$ & $\sigma_{E\gamma}=6.25$           & $\sigma_{E\gamma}=12.86$ \\

$m_{\gamma_i \gamma_j}-M_{\gamma \gamma}$, MeV/c$^2$   & 
$M_{\gamma\gamma}=130.72$ & $\sigma_{m_{\gamma\gamma}}=22.05$ & $\sigma_{m_{\gamma\gamma}}=4.5$ \\

$\theta^{exp}_{\pi^0 \gamma_i}-\theta_{\pi^0 \gamma}$, deg & 
$\theta_{\pi^0\gamma}=107.53$ & $\sigma_{\theta_{\pi^0\gamma}}=19.52$ & $\sigma_{\theta_{\pi^0\gamma}}=17.52$ \\

$E^{calc0}_{\pi^0(\gamma_i\gamma_j)}-E^{calc1}_{\pi^0(\gamma_i\gamma_j)}-\Delta E_{\pi^0}$, MeV & 
$\Delta E_{\pi^0}=-2.58$  & $\sigma_{E_{\pi^0}}=12.12$ & $\sigma_{E_{\pi^0}}=9.31$ \\

$\theta^{exp}_{\pi^+ \gamma_{6-i-j}}-\theta_{\pi^+ \gamma}$, deg & 
$\theta_{\pi^+ \gamma}=129.18$ & $\sigma_{\theta_{\pi^+\gamma}}=32.60$ & $\sigma_{\theta_{\pi^+\gamma}}=15.10$ \\

$\theta^{exp}_{\pi^+ \pi^0_{i,j}}-\theta_{\pi^+ \pi^0}$, deg & 
$\theta_{\pi^+ \pi^0}=150.04$  & $\sigma_{\theta_{\pi^+\pi^0}}=24.43$  & $\sigma_{\theta_{\pi^+\pi^0}}=9.67$ \\
\hline
\end{tabular}
\label{tab:Q_param}
\end{center}
\end{table*}

Since the distribution of the expressions in parentheses of each term's 
numerator in (2) is asymmetrical, the root mean square ($\sigma$) 
corresponding to each term has two values, one corresponding to the 
expression being $<0$, and the other to the expression being $>0$. 
At this stage the variables of $Q_{i,j}$ have been optimized to maximize 
the pairing efficiency of the DE component using the MC simulation, 
which we obtained to be 88\%. For the IB component we obtained 70\% 
efficiency. Sub-indexes of the minimum $Q_{i,j}$ are accepted to be the 
indexes of the two photons, which come from the $\pi^0$. We number all three 
photons in the following way: the most energetic photon from the $\pi^0$ 
decay is ascribed the number 1, the lower energetic photon is labeled by 2.
We ascribe number 3 to the third unpaired photon.

\begin{figure}[h]
\begin{center}
\includegraphics[width=12cm]{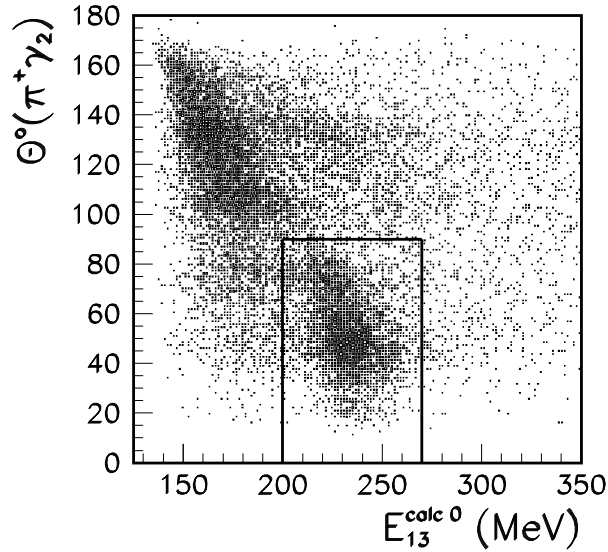}
\end{center}
\caption{Scatterplot of $\theta_{\pi^+ \gamma 2}$ vs $E^{calc 0}_{1 3}$. Here,
$\theta_{\pi^+ \gamma 2}$ is the opening angle between ${\pi^+}$ and
$\gamma 2$, and $E^{calc 0}_{1 3}$ is the parameter $E^{calc 0}_{\pi^0(\gamma_i\gamma_j)}$
with $i=1$, $j=3$. The detached region is a region of incorrect photons pairing.}
\label{fig:ppig13_ep0iv13}
\end{figure}

At the second stage we represent the events in a scatterplot as shown in
Fig.~\ref{fig:ppig13_ep0iv13}. 

As the MC simulation shows the region  in the rectangular box is a region of 
photons mispairing. 
Namely, for these events the $\gamma_2$ doesn't really come from the $\pi^0$ decay 
but is the unpaired one. Consequently the $\gamma_3$ is actually the lower energy 
photon from the $\pi^0$ decay. In order to determine the true combination of the 
photons in the detached region as well, we renumber the events in this region as 
follows: we ascribe number two to the third photon and number three to the second 
one. Such simple cross-renumbering of events allowed us to enhance the pairing 
efficiency of the IB component up to 93\%, although the DE component was 
reduced down to 81\%. Nevertheless, on the whole we increased the pairing
efficiency considerably, and this in turn reduced the systematic error
associated with mispairing. For comparison, in the previous analysis~\cite{Shim2} 
the pairing efficiencies were 85\% and 79\% for IB and DE components, respectively.

\begin{figure}[h]
\begin{center}
\includegraphics[width=12cm]{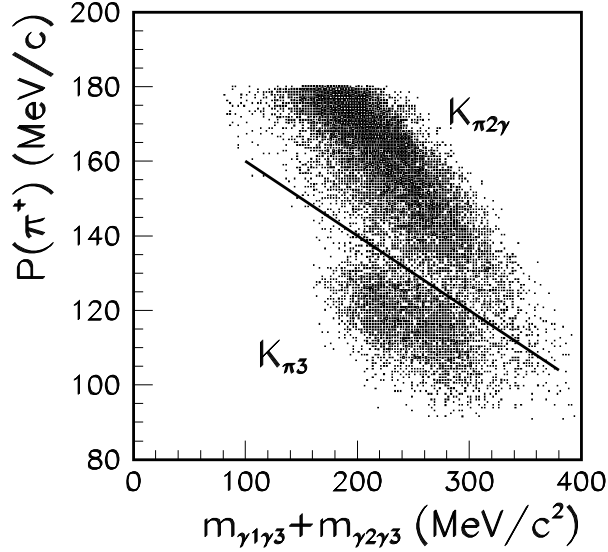}
\end{center}
\caption{Scatterplot of the $\pi^+$ momentum versus the sum of "invariant masses" 
comprised of $\gamma_1$, $\gamma_3$ and of $\gamma_2$, $\gamma_3$. The shown line 
is a cut condition to remove the $K_{\pi3}$.}
\label{fig:ppi_mgg1_mgg2}
\end{figure}

The $K_{\pi3}$ background was rejected by a 2-dimensional cut as shown in
Fig.~\ref{fig:ppi_mgg1_mgg2}.
Such a cut not only allowed us to
successfully reject this mode but also 
saved good $K^+ \rightarrow \pi^+ \pi^0 \gamma$ events in the $\pi^+$ low momentum region.

The total number of good $K^+ \rightarrow \pi^+ \pi^0 \gamma$ events in the $\pi^+$ momentum region of
115 to 180 MeV/c, obtained after applying the above selection criteria, is 
10,154, which is about 2.3 times higher than in the previous analysis~\cite{Shim2}. 
The background from $K_{\pi3}$ decay has been estimated to be less than 
0.2\% by the simulation. In the previous analysis the contribution of this 
mode was about 1.2\%.

\begin{figure}[h]
\begin{center}
\includegraphics[width=11cm]{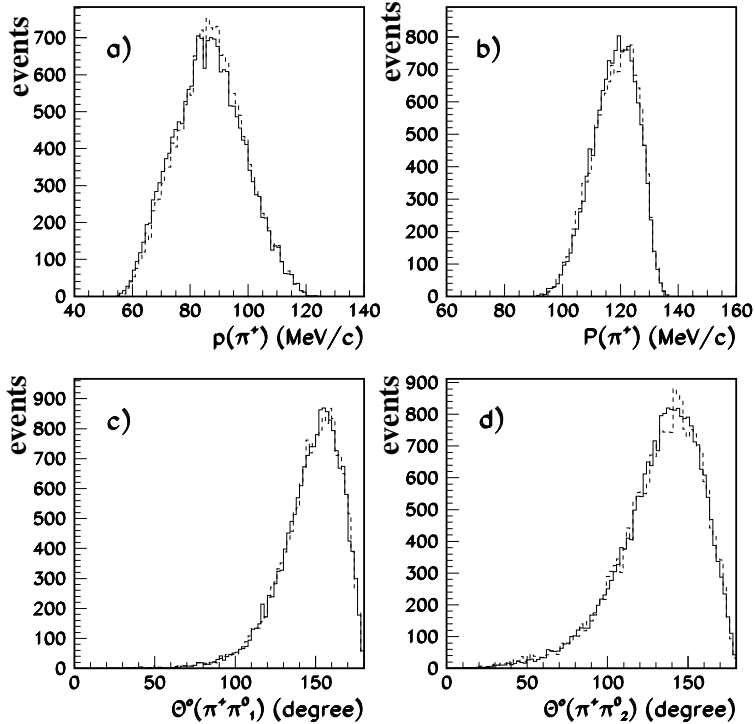}
\end{center}
\caption{$K_{\pi 3}$ spectra: 
a) - $\pi^+$ momentum before correction for energy losses in the target; 
b) - $\pi^+$ momentum after  correction for energy losses in the target.
c) - opening angle between $\pi^+$ and $\pi^0_1$; 
d) - opening angle between $\pi^+$ and $\pi^0_2$. 
The solid lines are the experimental data and the dashed lines are the MC predictions.}
\label{fig:kpi3_spectra1}
\end{figure}

\begin{figure}[h]
\begin{center}
\includegraphics[width=11cm]{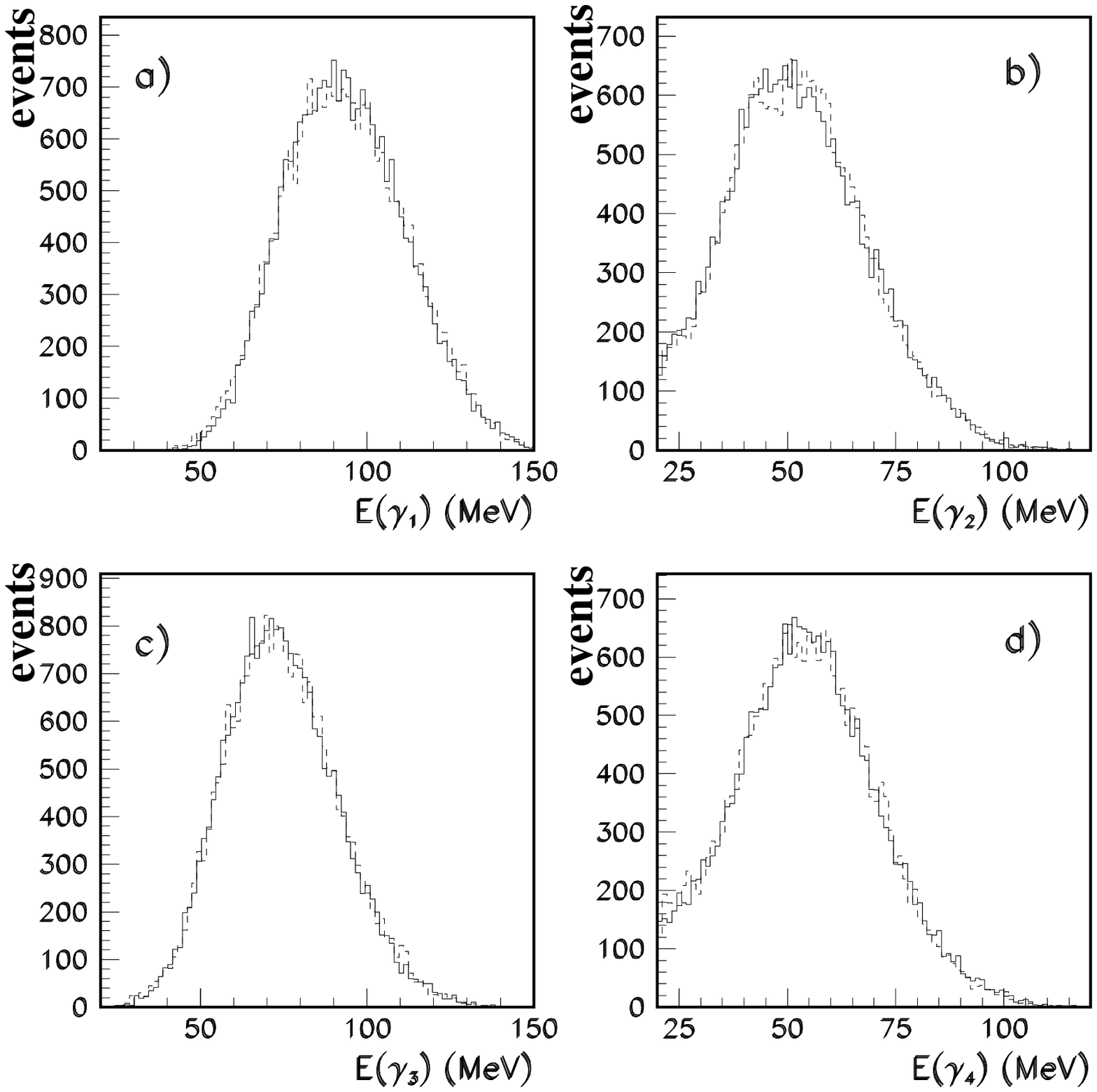}
\end{center}
\caption{Spectra of $K_{\pi 3}$:
a) - energy of $E_{\gamma 1}$;
b) - energy of $E_{\gamma 2}$;
c) - energy of $E_{\gamma 3}$;
d) - energy of $E_{\gamma 4}$.
The solid lines are the experimental data and the dashed lines are the MC predictions.}
\label{fig:kpi3_spectra2}
\end{figure}

A MC simulation of the $K_{\pi 3}$ decay was used to test the simulation of
the experiment. The matrix element of $K_{\pi 3}$ decay for this simulation 
was taken from the PDG~\cite{PDG}. In the simulation, the form factors of 
the $K_{\pi 3}$ matrix element were taken to be the world average quoted by 
the PDG~\cite{PDG}. In this analysis we selected events with 4 photon 
clusters in the Cs(Tl) using the following conditions, 
$P_{\pi^+}<135$~MeV/c,
$14000<(M_{\pi^+}^{tof})^2<25000$ (MeV/c$^2$)$^2$,
$\Delta p<80$~MeV/c,
$380<M_K<520$~MeV/c$^2$,
$80 <E_{\pi^0_1}<200$~MeV,
$60 <E_{\pi^0_2}<190$~MeV,
$60 <m_{(\gamma \gamma)1}<150$~MeV/c$^2$,
$55 <m_{(\gamma \gamma)2}<150$~MeV/c$^2$.
Various $K_{\pi 3}$ spectra are shown in Figs.~\ref{fig:kpi3_spectra1},
\ref{fig:kpi3_spectra2},
and we see excellent agreement between the experimental data and the MC simulation.

\begin{figure}[h]
\begin{center}
\includegraphics[width=11cm]{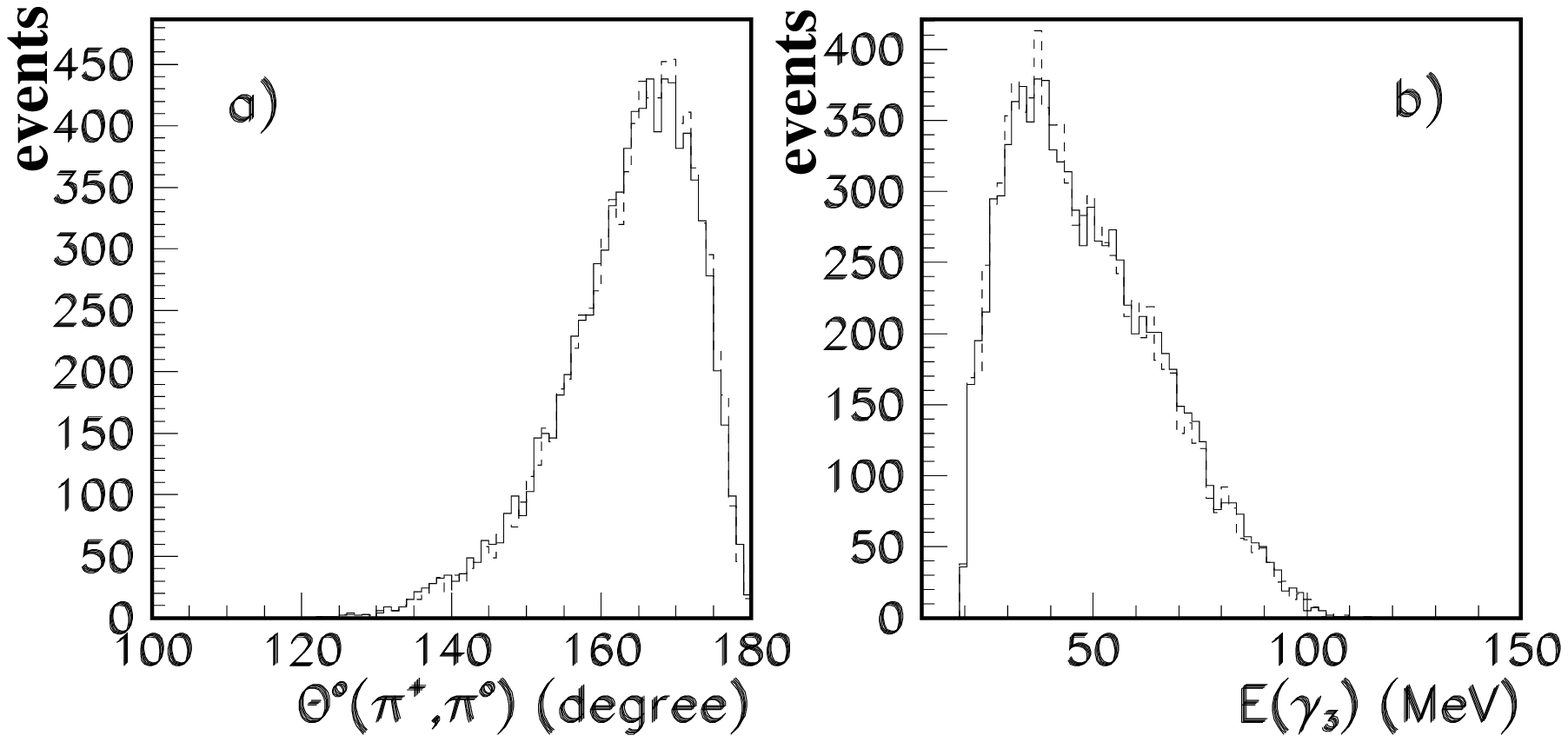}
\end{center}
\caption{Spectra of $K^+ \rightarrow \pi^+ \pi^0 \gamma$ with an additional cut of $W<0.5$. 
a) opening angle between the $\pi^+$ and the $\pi^0$;
b) energy of $\gamma 3$. The solid lines are the experimental data and 
the dashed lines are the IB component from the Monte Carlo predictions.}
\label{fig:kpi2g_spectra}
\end{figure}

A MC simulation of the $K_{\pi2\gamma}$ decay was carried out for both the 
IB and the DE components of the $K_{\pi2\gamma}$ decay, using the matrix 
elements given in~\cite{ecker2}.
The detector acceptances for the IB and the DE components have been
obtained to be
$\Omega(IB)=0.566\times 10^{-3}$ and $\Omega(DE)=1.659\times 10^{-3}$.
Figure~\ref{fig:kpi2g_spectra} 
shows the $K^+\rightarrow\pi^+\pi^0\gamma$ spectra, which have been obtained 
after imposing an additional cut of $W<0.5$ in order to remove the DE component. 
Again we find very good agreement between the experiment and the MC simulation.

\section{Results}

\begin{figure}[h]
\begin{center}
\includegraphics[width=9cm]{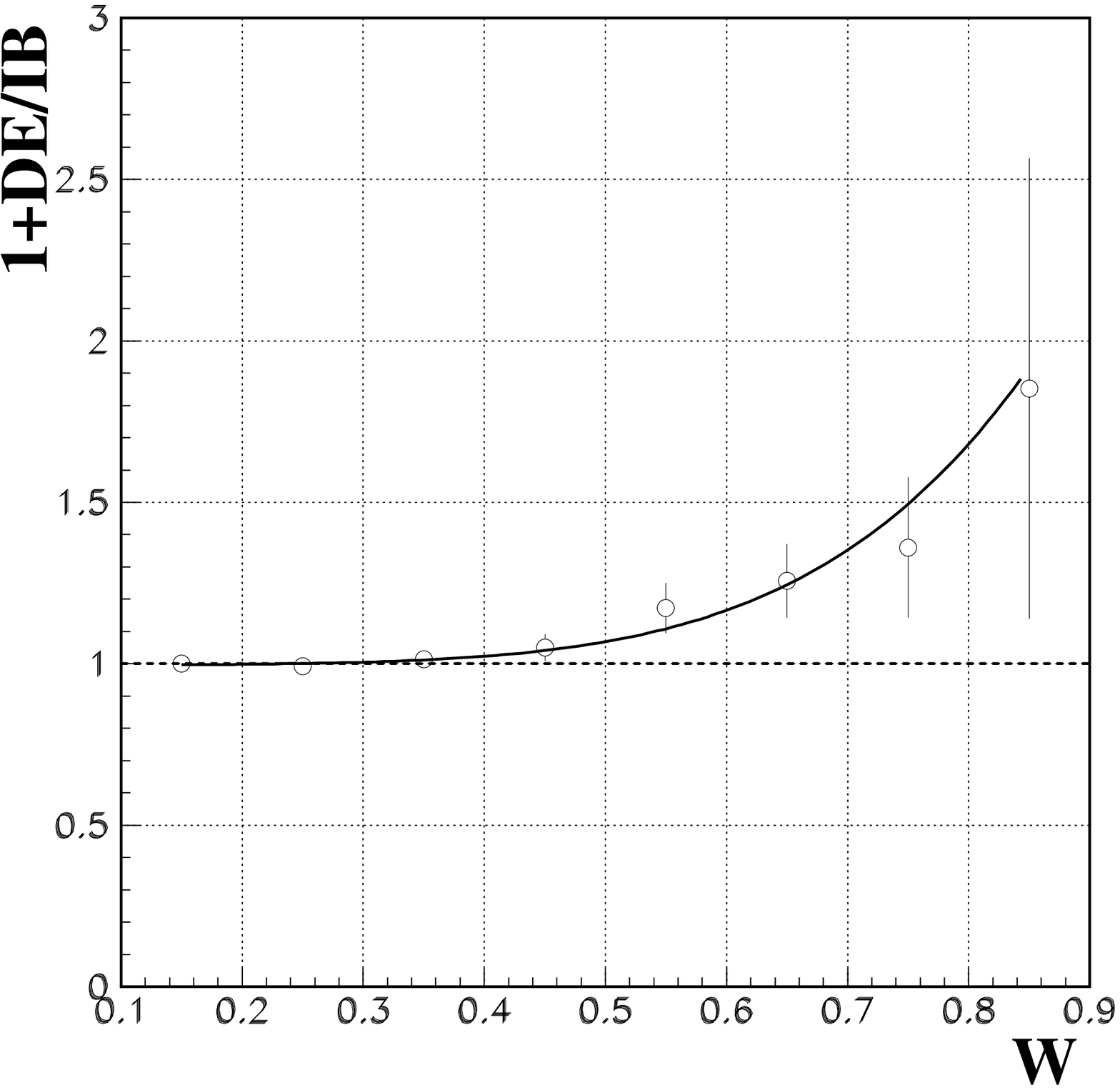}
\end{center}
\caption{W spectrum of the observed $K^+ \rightarrow \pi^+ \pi^0 \gamma$ 
events normalized to the IB component obtained from the Monte Carlo simulation. 
The solid curve shows the best fit of the W spectrum to a sum of the IB and DE 
components.}
\label{fig:wfit}
\end{figure}

The fraction of the DE component in the selected $K^+ \rightarrow \pi^+ \pi^0 \gamma$ events has been
obtained by fitting the experimental data distribution of 
$\theta_{\pi^+\pi^0}$, $E_{\gamma}$ and $W$ in 3-D space with a sum of IB 
and DE components obtained from the MC simulation. The Maximum log-likelihood 
method has been used for fitting the $\chi^2$ defined as follows
\begin{equation}
\label{eq:chi2}
\chi^2=2\sum[y_i-n_i+n_i\ln(n_i/y_i)],
\end{equation}
where $y_i=A\times(y_i^{IB}+\alpha y_i^{DE}).$ Here, $n_i$ is the number of 
real data in the $i$-th bin of the histogram, and $y_i^{IB}$, $y_i^{DE}$ are 
numbers of MC data of IB and DE components in the same $i$-th bin of the 
histogram, respectively.  The variable $\alpha$ is the fraction of the DE 
component normalized to the IB one and  $A$ is a normalizing coefficient. The 
MINUIT package was used to obtain the minimum of $\chi^2$. The error values 
correspond to shifts of the variables, which increased the $\chi^2$ by 1. 
The value of $\alpha$ was obtained to be $0.026 \pm 0.006$ with a 
$\chi^2/ndf$=1.15. The experimental spectrum of $W$ normalized to the pure 
IB component is shown in Fig.~\ref{fig:wfit}.
The obtained branching ratio of the DE component in the $\pi^+$ kinetic
energy region of 55 to 90 MeV is
$
Br(DE) =[3.8 \pm 0.8(stat)] \times 10^{-6}.
$

The major systematic errors come from the uncertainty of the CsI(Tl)
calorimeter calibration, $\gamma$ mispairing effect, accidental background, 
the spectrometer field uncertainty, and also from the uncertainty of the 
kaon stopping $x$, $y$ position.

The uncertainty due to the $\gamma$ calibration of the CsI(Tl) calorimeter
was estimated by changing the value of the DE branching ratio by the
maximum shift of the calibration coefficients of the CsI(Tl) calorimeter. 
The corresponding shift of the DE branching ratio obtained in this way was 
$\pm 0.5 \times 10^{-6}$.
The $\gamma$ mispairing effect was studied by changing the pairing
efficiency of the photons along with a change of the selection cuts. 
The shift of the DE branching ratio as a result of such changes was 
$\pm 0.4 \times 10^{-6}$.
The uncertainty from the accidental background, mainly associated with the
$K_{\mu 3}$ decay, was estimated by varying the TDC gate widths, and was
found to be less than 7\% of the DE branching ratio value, that is 
$\pm 0.3 \times 10^{-6}$.
The error due to the spectrometer field uncertainty ($\pm 0.1 \times 10^{-6}$) 
was evaluated by shifting the field strength by $\pm 0.2\%$.
The error due to the uncertainty in the reconstruction of the kaon stopping
$x$, $y$ position was studied by a shift of the kaon stopping point 
along the charged pion track by one fiber which corresponds to 5~mm. 
The resulting variation of the DE branching ratio was $\pm 0.2 \times 10^{-6}$.
All the uncertainties are summarized in Table~\ref{tab:syst.errors}.
\begin{table}[h]
\caption{Summary of the systematic errors.}
\label{tab:syst.errors}
\begin{center}
\begin{tabular}{|c|c|}
\hline
                                    &  UNCERTAINTY,                  \\
 ERROR \ SOURCE                     &  $\Delta Br(DE)\times 10^{6}$  \\
\hline
 $\gamma$ \ calibration \           &  0.5  \\ 
 $\gamma$ \ mispairing \            &  0.4  \\
 Accidental \ background \          &  0.3  \\
 Spectrometer field uncertainty      &  0.1  \\
 Kaon stopping $x$, $y$ coordinates &  0.2  \\
\hline
Total \ systematic \ error          &  0.7  \\
\hline
\end{tabular}
\end{center}
\end{table}

\section{Conclusion}

We have obtained an improved result for the branching ratio of direct photon 
emission in the $K^+ \rightarrow \pi^+ \pi^0 \gamma$ decay from the analysis 
of the E470 data. We have extracted 10154 $K^+ \rightarrow \pi^+ \pi^0 \gamma$ 
events in the $\pi^+$ momentum region of 115 to 180 Mev/c, which is about 2.3 
times more than in the previous analysis~\cite{Shim2}. In addition to a 
reduction of the statistical error we have also reduced the systematic error 
by increasing the pairing efficiency of the photons. The optimization of the 
selection cuts allowed us to increase the total number of the extracted 
$K^+\rightarrow\pi^+\pi^0\gamma$ events, and to obtain more sensitivity to the 
DE component part of the $K^+ \rightarrow \pi^+ \pi^0 \gamma$ spectrum. In this 
analysis we have provided more efficient rejection of the $K_{\pi 3}$ 
background, which was estimated by MC simulation to be less than 0.2\% as 
compared to 1.2\% in the  previous analysis. 

The branching ratio for the DE component in the $\pi^+$ kinetic energy 
region 55 to 90~MeV has been determined to be
$$
Br(DE)=[3.8 \pm 0.8(stat) \pm 0.7(syst)] \times 10^{-6},
$$
which is consistent with the results of the three last experiments~\cite{Adler,Shim2,Uvar}.
The good agreement of this result with the theoretical prediction for the branching 
ratio of the DE component, $Br(DE)=3.5 \times 10^{-6}$~\cite{ecker1,bijnen}, 
supports the hypothesis that the dominant contribution to direct photon emission 
is due to the pure magnetic transition given by the reducible anomalous amplitude.
It should also be stressed that the result has been obtained with the assumption 
that the INT component is zero.

~\\
\emph{Acknowledgments.}
{\small This work has been supported in Japan by a Grant-in-Aid from the Ministry
of Education, Science, Sports and Culture, and by JSPS; in Russia by the 
Ministry of Education and Science; in Canada by NSERC and IPP, and by the 
TRIUMF infrastructure support provided under its NRC contribution. The 
authors gratefully acknowledge the excellent support received from the KEK 
staff.}

\end{document}